# A Criticism of the Current Security, Privacy and Accountability Issues in Electronic Health Records

Adebayo Omotosho
Department of Computer Science, College of ICT,
Bells University of Technology, Nigeria

Justice Emuoyibofarhe
Department of Computer Science and Engineering,
Faculty of Engineering and Technology, Ladoke
Akintola University of Technology, Nigeria

## ABSTRACT
Cryptography has been widely accepted for security and partly for privacy control as discovered from past works. However, many of these works did not provide a way to manage cryptographic keys effectively especially in EHR applications, as this is the Achilles heel of cryptographic techniques currently proposed. The issue of accountability for legitimate users also has not been so popular and only a few considered it in EHR. Unless a different approach is used, the reliant on cryptography and password or escrow based system for key management will impede *trust* of the system and hence its acceptability. Also users with right access should also be monitored without affecting the clinician workflow. This paper presents a detailed review of some selected recent approaches to ensuring security, privacy and accountability in EHR and gaps for future research were also identified.

## 1. INTRODUCTION
Though security and privacy are strongly related but the two concepts are obviously differing. Privacy is the right of an individual to determine for themselves when, how and to what extent information about them is shared or transfer to others while security on the other hand defines the extent to which personal information access is restricted only to authorized personnel [9]. Unauthorized transfer and sharing of sensitive health data could result in several unwanted usage which could result in, for example, unwarrantable discrimination by employers and so on. Also, due to the fact that there are some organizations such as government, pharmaceutical companies, employers, laboratories and researchers may have justifiable reasons to access patients' health information, health care personnel could accidentally or intentionally abuse record access privileges. Many a time, privacy is also breached by the unavoidable systemic identification that takes place throughout the electronic health infrastructure and with idea of central parties and technologies that observed all patient and healthcare service provider actions.

Critical to the maintenance of trust with the health care providers and acceptance of EHR is the patients' perception of security and privacy of health records. In an age of identity theft and data snooping, the health care industry has become one of the most sought after domain by cyber criminals as the transition from paper based health systems to electronic health records (EHRs) has given data thieves compelling reasons to attempt cracking of hospital networks due to the value of medical data it contained. Electronic medical records therefore are vulnerable to potential abuses, losses, leakages and threats [12]. In recent years, hundreds of thousands of patients' health information has been made liable to danger due to security lapses at hospital and government agencies [6]. A patient based survey on EHR carried out by [11] showed that majority of patients are willing to embrace EHR transition but are seriously concerned about the privacy and security of their health records. The veracity and completeness of stored data can be deteriorated if these perceived risks are not controlled as in some cases patients may resort to falsifying information as an alternative, in order to preserve their privacy. With most culprits making enormous profits from data theft and misuse at the detriment of the patients' privacy, EHR need to be more protected from illegal access and usage. Information security ensures the protection of personally identifiable information in records managed with EHR from compromise, unauthorized access, use, disclosure, modification, destruction, disruption or other situations where unauthorized persons have access or potential access to such information for unauthorized purposes.

Also, putting into consideration that today's ever-increasing requirements for high security standards, in order to secure all kind of important information, the science of cryptography has become even more important. However, in generic cryptographic systems user authentication is still possession based [42]. This implies that the possession of a cryptographic key needed to authenticate a user. Generally, in most cryptographic key management systems these keys are released by presenting a password (or PIN), determined by the user, to the system. This connotes that the cryptographic key is just as secure as the password which is used to release it and these passwords are often chosen weakly as is all too well known. Additionally, a physical token such as a smartcard can be lost or stolen.

## 1.1 Research Rationale
Electronic health record (EHR) systems are expected to ease the process of sharing health information among health care providers which in turn improve quality of health care delivery. EHR promises monolithic benefits in terms of saving cost by digitizing and centrally providing medical data [37]. They serve as the repository for valuable health information which is assets to both the health care provider and the data criminals. Their continuing misuse and fear of usage abuse have however posed an unnerving trust challenge because the patients' personal data and anamnesis stored and transmitted via this system can be susceptible to various arrays of attacks such as the medical identity theft. Also, for different reasons, individuals may not wish for personal data such as their religion, sexual orientation, HIV/AIDS diagnosis and treatment, reproductive health, substance abuse, mental health, genetic conditions, and sexual assault to be revealed as this may be to avoid irreversible personal embarrassment, discrimination or damage to ones professional reputation [36]. An insecure EHR system could also results in endangering patient healthcare, inadequate quality of health service.

The case of Insider threat is a relatively new and not much has been written on it, an insider attacks from those who have legitimate access to the EHR system and lack of access control mechanism have contributed largely to this mayhem [32].

Today, cryptography has been widely accepted for security and privacy control for example in, [7], [1], [41], [29], [27], [16] and many more. However, these works did not provide a way to manage cryptographic keys effectively as this is the Achilles heel of cryptographic techniques currently proposed for EHR systems. Accountability also has been considered in limited works relating to EHR such as in [41], [29] and [14]. However, these authors' accountability modes suffered the major flaws of flexibility and physician privacy protection which may affect physician adoption of such systems. Unless a different approach is used, the reliant on cryptography and password based system for key management will impede *trust* of the system and hence its acceptability.

## 1.2 Significance of Study

Research has shown that insider threats are more difficult to address than external threats because individuals perpetrating the crime are authorized personnel, friends and co-workers which makes it difficult to identify the criminal. Recently, Information and Communication Technologies (ICT) have transformed patients' role from the traditional passive recipient of healthcare services into a more active role in which patient have more understanding of their health record and are empowered with the ability to make choices and be involved in decision making process [13]. This has given rise to the challenges of what degree of freedom to be given to issuers, data subjects or consumers in managing EHR without hindering clinicians' workflow and compromising security and privacy.

Electronic health record overcomes most of the drawbacks of the conventional paper records such as errors arising from illegibility and it supports easy backing up of heath data which prevent data loss in contrast to the paper approach [5, 38]. Health information managed by EHR should be accessible, available and remain unchanged at all times therefore both the accountability and integrity of such information need to be verifiable. If the privacy and security mechanisms tailored toward controlling access to medical data are not too cumbersome and socially uncomfortable for both the patients and physicians many of the benefits of accessibility, timeliness and quality health care delivery would be materialized.

## 2. RELATED WORK

Recently, the subjects of security and privacy in electronic form of health record have generated a lot of controversies in the adoption of EHR. The question of access rights to data, how and when data is stored, security of data transfer, data analysis rights and the governing policies posed an unending challenge in electronic health acceptance and remained research questions that need to be answered. For EHR to be safer and widely adopted there is currently a need for the development of such a system that meets today's EHR systems requirements. Several pilot projects and models have explored secure storage and access to health records.

## 2.1 Popular Approaches to Controlling Privacy and Security

On the use of access control as a methodology for enforcing privacy of patients in EHR, [24] proposed a security methodology that uses Role Based Access Control as a means to ameliorate most of the privacy and security related issues associated with web based electronic health care record. Rather than giving over control of EHR to all of General Practitioners in a heath setting, privacy is achieved by associating roles with each individual who might have a need to access information, with each roles defining the set of privileges and operations an individual assuming that role may perform. The overall system adopted concept of the roles, the authorisation management and the roles hierarchy and the inheritance. The authors also supported the idea that the provision of security method for communication over insecure public internet requires the use of cryptographic and authentication techniques. However, their proposed Role Based Access Control model merely shows an access control matrix which manages objects a specified role could access and did not consider the data subject control over privacy.

[24] proposed approach may result in stalemate in real application when trying to achieve security and privacy because EHR privacy could also preferably be jointly managed by both the data subject and the health care professional according to [9], who also supported cryptographic technique for security measure. [9] discreetly criticised the over reliant and over stretching of the Public Key Infrastructure (PKI) in access control. He proposed SPACER- **S**ecure and **P**rivacy-enhanced **A**ccess **C**ontrol for **E**-health **R**ecords which allows EHR to be stored on smartcards, mobile phones and other portable devices while enforcing secure management of EHR by both the patient and the general practitioner as partial owners. The SPACER approach may virtually reduce security and privacy risk of patients without affecting the workflow of the health care. However, the use of smartcards proposed to be used by patient for access control may not be too reliable as smartcards are issued once but can be misplaced.

[20] approach to protecting health information system theorized access control requirement and argued that access control system like the one defined in [24] cannot adequately consider real world methods for roles due to the complexity in defining constraints. The authors noted that despite the sensitivity of the data and the rising threat, not much attention has been given to the complexities of real-world access constraints. The authors like [9], stressed the much hype encryption techniques has been receiving and rounded up by describing a two-level mechanism that can fulfill minimum access requirement criteria. The theory was not implemented and besides it would be difficult to provide security to today's system without encryption.

The security and privacy implications that may arise when integrating new technologies into the traditional health care system were uniquely identified in [31]. The authors stated that the issues of data access, data storage and analysis are however not peculiar to the medical field alone and that similar problems have been seriously considered in other areas like the financial services and internet shopping and that there exist technical solutions that can be applied to EHRs in order to address these similar issues in multi-user settings. Similar to existing work, at the end of their study, they also suggested future implementation of role based access control, encryption and authentication mechanism as likely solutions.

[2] examined patient's privacy and data security risks inherent in the transitioning from paper health record storage to the electronic approach. Again, contrary to the work of [24], it was shown that no single of discretionary, mandatory, or role-based access control techniques in isolation could effectively meet the privacy and security requisites of an EHR. However, similar to [9], the authors criticised the current focus on PKI which is by design primarily for securing data in transit where neither the data subject nor the receiver access is safe. The authors proposed a hybridized access control mechanism that securely conglomerated the three traditional security models for access controls and formulated a new combined access control protocol. Joint management of EHR was also promoted; however, the model did not considered cryptography as an option to achieving security and privacy, so until such system is implemented it will be difficult to adjudge how the hybridized protocol will thrive in reality.

Most research endorsed encryption as a near definite solution to security but, the concern over the sizes of medical data, especially medical images, being encrypted without consuming time appears infeasible as this could creates a shortcoming in a system that relies wholly on encryption as a security mechanism for EHR. Adoption of encryption is growing geometrically, as a matter of fact an example of a multi-layer encryption was proposed by [18]. [18] presented an over-encryption technique in the management of access control evolution on outsourced data. The work adopted a two-layer encryption on data; one by the data owner and the other by the server. To handle the accelerative data volume, [28] approached the problem of privacy and security of e-health from the perspective of pseudonymization, **P**seudonymization of **I**nformation for **P**rivacy in **e**-Health (PIPE) was introduced which is a complete patient-centric security approach that integrates primary and secondary health data usage. In their system, instead of encrypting actual medical data, patients' identification tags are transformed and stored as pseudonyms which are generated using symmetric or asymmetric encryption algorithms. Unlike most pseudonym based system, PIPE does not rely on a patient list in order to relate patients' identity with medical data. A patient uses smartcard containing a secret key to grant or revoke access to their medical record as they are given full control. Though smartcards may be lost or stolen, the system provides a data (secret key) recovery mechanism through super administrator of the system using RBAC. This however does not guarantee privacy as smartcard PINs could be learnt or stolen and then used to access patients EHR. Another challenge in this work is the granting of the full EHR access to patients who has no knowledge of the information that would be needed for each medical practitioner for immediate and emergency treatment.

To further buttress the need for the enforcement of security and privacy policies for patients in an electronic health care setting, [7] argues that hierarchical encryption system and access control should be deployed. Similar to [28] a patient-centric security approach was proposed. [7] presented the concept and implementation of asymmetric and symmetric key Patient Controlled Encryption (PCE) whereby patients generate and store personal encryption keys, this way; if the host data centre be compromised the patients' privacy is protected as the server that stores the health information will never have access to keys given to the doctor and hence will be unable to decrypt any of the data. In PCE framework, patients use their decryption key to generate sub keys which will allow his/her delegates to access only a certain parts or portions of her record. This approach provides a high level of security to EHR by preventing unauthorized access and privacy breach of patients' records however, the system may not be practicable in the case of emergency because the patient fully controlled data encryption and access right to record parts, the time overhead in requesting access per patient may not be worthwhile. Also, the issue of key management by both patient and doctor may be tedious.

## 2.2 EHR, Accountability and Keys Availability

Hitherto, most of the focuses of EHR security applications have been directed to preventing external threats from accessing health information with techniques that heavily relied on cryptography technologies without serious consideration of malicious insider threats and a reliable method for accountability reporting. Due to the progressive rising of insider attacks on organizations, [8] proposed the technique of baiting inside attackers using decoy documents to confuse malicious users, this trap-based defense technique automatically hides the actual information amongst misleading information which is saved as a file system document. When a decoy document is opened, the information about where and when it was accessed is transmitted to the monitoring server. Though the work did not focus on health data but it provides an insight into detecting a malicious attempt on sensitive information. The procedure proposed as well may not be adequate to handle EHR data since medical data are not just an array of textual documents.

Captivatingly, [1] carried out an extensive review on the issue of securing electronic health data transmissions over insecure communication channels and similar to other literatures, the authors also affirmed that encryption methods are efficient ways to protect data. [1] however reported from their extensive study that most of the secure systems and architectures proposed so far all suffers a major flaw by not discussing the patients' right and how the system can pinpoint the person who broadcast medical records for accountability responsibility.

Meanwhile, at [41] stated that the effective protection of patient data and privacy cannot be achieved without the patient being in control of their health information. Going by this, the authors proposed cryptography based secure EHR system for ensuring the protection of patient privacy in the case of emergency since patient lock and unlock access to health record. In addition, the proposed system restricts protected health information access to only authorized physicians, who can be traced and held accountable if the accessed health data is found improperly disclosed. This work attempted to answer [1] by providing a form of accountability, although only for emergency cases but the work assumes that the patient are not completely out as they are responsible for prior delegating of access to certain part of their EHRs. Also, the inpatient grants transitive emergency access via a P-device through their family member to the doctor. The cryptographic keys are known to the family members alone, nevertheless no attempt was made on how these keys could also be managed.

[14] also identified that emergency access represents one of the easiest methods to access unauthorized data because the malicious users needs only to provide a plausible reason for

access. The proposed system largely depends on the cooperation between the patient and medical provider, and between medical providers themselves to achieve its main goal. However, if the providers did not take time to mark the emergency data and just marked everything then the system becomes useless, it would become a burden on the EHR system and becomes a major security hole in the system.

Rather than using a patient mobile device for emergency access, recently, [17] proposed an alternative approach which uses biometric identification to access a central health record database during emergency. The method used was to provide the technicians with a mobile system through which they gain access to necessary attributes of patients EHR using the patients fingerprint during emergency.

[29] also argue that encryption-based protections, including [7], are not adequate on their own to guarantee patient awareness and control of health record. Despite the fact that recent researches have revolved around the magnitude of data subject control over EHR access, there is still a need for a method for accountability use and update in a patient centric approach as demanded in [1]. The demand for who should be answerable in EHR was as well included in [29] who proposed a cryptographic based mediation protocol for a patients monitoring agent that ascertains and logs the access of both health record issuers (e.g. medical practitioners) and consumers (e.g. government) whenever they use, share or update health data stored in an EHR system. The principal goal of the system is that patients would always be aware of any usage access to their health data which helps to immediately identify malicious attacks, sharing or threat especially after health records are released. Unfortunately, this work did not consider the information flow challenge as well as those from inside attacks which should be of great concern as this contributes largely to health record abuses. Likewise the protection of issuer's identity is not addressed in this work and could potentially result in privacy violation.

Immediate actions must be taken to resolve all the technical issues, which will surely increase the adoption of EHR [35]. Overall, it is undeniable that in their own way, encryptions appear to be providing good security except for the key storage problem associated with them. Key management remains a serious issue in all cryptographic based systems. In this view, [27] though did not considered accountability aspect of EHR but attempt to improve on the existing scheme; the framework is also similar to [7] but in contrast their work they divided users in the system into multiple security domains in order to reduce the key management complexity for data owners and users. Attribute based encryption technique was used to encrypt patient health data. The work however did not discuss how the keys will be securely stored and managed by either party because careless key leakage voids encryption. [16] work was also an improvement over [7], still another effort to reduce key management issues but also implemented attribute based encryption. The challenge of the work still lied in lot of keys being stored in plain form by a central authority.

## 2.3 Cryptography Key Issues, Biometrics Keys and Protection Approaches

In order to properly prevent insiders' threats in EHR systems and to continuing the wide adoption of cryptography some major challenges in disguised often overlooked have to be addressed. Since all cryptographic algorithms rely on known keys, some works considered improving the strength of cryptographic algorithm with *biometric* based keys protection, generation or binding for several reasons. Conventional cryptographic keys used for encryption and decryption are long and random, hence cannot be memorized. This has led to storing the cryptographic key in some other position and release it based on some alternative authentication like password which could as well be guessed or stolen. Cipher keys may be illegally shared and this would void non-repudiation, biometric can be used to protect or generate cryptographic algorithm keys which could help to alleviate the problem of key managements in the current cryptographic encryption implementation. Without proper key protection EHR insiders threats could be aggravated. Bio-cryptography, is however very challenging, all the same it is expected to provide huge benefits over cryptography.

Despite the promising combined advantages of uniqueness and hard security, if biometric templates themselves are not protected, bio crypto technology will fail to halt insider threats. Therefore, the protection of biometric template in bio cryptosystem is of paramount importance in order to maximizing the joint benefits of biometric and cryptography. One way to protect biometric template used with cryptographic key was proposed by [15] which was to encrypt biometric templates or images stored in a database using conventional cryptographic methods as this would improve the level of the system's security, since an intruder must gain access to the encryption keys first before an attack can be launched. This method however did not solve the most privacy issues associated with a large database since the keys and the biometric data are controlled by an administrator. Apparently, the expected role of biometric in the traditional cryptosystem is to improve key management. [39] proposed another method which was based on the ground that biometrics systems either yield a one bit Yes or No information, if a Yes response is produced because a user is confirmed genuine then system unlocks a password or a key. The security of users' keys is ensured by storing them in a secure location. This scheme is still prone to the security vulnerabilities since the biometric system and the application are connected via one bit only.

[3] used data derived directly from a biometric image to generate cryptographic keys. Since the quality of biometric data relies considerably on individuals physiological traits and is also strongly influenced by environmental factors; it is therefore characterized by inaccuracies. Generating cryptographic keys directly from biometric data is challenging since biometric data are not always the same to ensure consistent key are being generated. [43] suggested a method that involves hiding the cipher key within the biometric enrolment template itself through a secret bit-replacement algorithm. If the user is successfully authenticated, the algorithm extracts the key bits from the appropriate place and releases the key. This is a very good scheme but biometric templates are very fragile any little modification to the original image could void the existence of the original template because this may likely happen during the bit replacement process. In another sequel, using [25] fuzzy vault key binding approach, [44] presented the results of a fuzzy vault implementation using fingerprint minutiae data. The experiment result showed that the vault performs with reasonable accuracy. The authors also affirm that the 128-bit AES keys can be feasibly secured using their proposed architecture. Multiple fingerprint data are captured per user to ensure higher accuracy however, this technique is not optimal.

Implementing the [26] fuzzy commitment approach to biometric template and key protection is considered very difficult. However, an insight into the practical use of fuzzy commitment was demonstrated by [23] who applied their own version of the fuzzy commitment scheme to iris codes. The system was tested with 700 iris images reaching a success rate of 99.5%. In addition, a False Rejection Rate of 0.47% and a zero False Acceptance Rate was recorded. These are very remarkable results which were not achieved until then, especially with iris scan because of the complicated engineering process of generating usable iris codes. However, the length of the keys used in the simulation was not provided as well as the systems response time.

[10] argued that biometric based encryption technologies have enormous potential to enhance privacy and security so far as keys are only accessible to legitimate users. Like fuzzy commitment, fuzzy vault is one of the most comprehensive mechanisms for secure biometric authentication and cryptographic key protection. It eliminates the key management problem as compared to other practical cryptosystems. In a key binding mode, [33] presented a fully automatic implementation of fuzzy vault scheme based on fingerprint minutiae. Due to fingerprint FAR the authors recommend future work to consider a way to reduce the likelihood of false acceptance. [47] also used fuzzy construct to store iris biometric template however, these templates are bound to a random key generated from the templates unlike in [33]. The authors further hardened the fuzzy vault through a password to provide an additional layer of security. However, unless a different transformation fuzzy technique is used the use of password may not be over secure as devised.

Multimodality bio-cryptosystems framework was proposed by [19] to be considered for future research in their security enhancement of cryptography, since each single biometric modality has its weakness. Multiple biometric mode system could reduce the errors found in a unimodal biometric system because an alternative approach, though not optimal, for increasing biometric system accuracy (reducing FAR/FRR) is to store multiple and redundant templates for each users. In continuation of their previous research, [34] again presented another work, but theoretical, on multi biometric template security using fingerprint and iris. The authors found that a multimodal biometric fuzzy vault renders a better performance and security compared to its counterpart unimodal biometric vault. [30] as well took another step into multimodal biometric template security by considering iris and retina template with password hardening. In another scenario, [40] performed a double AES algorithm encryption on the fuzzy vault itself. The authors also used multiple impressions of iris in order to provide higher accuracy rate. The main drawback of this work is the storage of the AES key since it will not be included in the vault.

[22] reviewed fuzzy vault biometric cryptosystem technology for protecting private keys and releasing them only when the legitimate users enter their biometric data. The authors also argued that fuzzy vault provides better security with iris and retina, because of their higher stability and template longevity as compared to other biometric traits. The major challenge in this work is how retina biometric with low ease of use could be deployed with iris in a real life application. [4] did not consider multi biometric based fuzzy vault but proposed iris based cryptography from which secret key is generated – bio cryptography key generation. Since iris is the most accurate biometric besides DNA, it should alone be able to provide

uniqueness. The authors perform an evaluation of the system to check key randomness. Symmetric algorithm, AES was deployed and information is encrypted and decrypted using the key. The evaluation result did not however prove that the key will always be the same which is a serious challenge for the symmetric AES used.

While bio-cryptography is still a growing field, creating cryptographic keys from biometric template directly is another possibility. [21] presented the concept of generating encryption keys directly from live biometric feature. Statistically generated synthetic biometric data were first used and then real biometrics (handwritten signature) in their proposed methodology. Expectedly, results from using synthetic features showed that under appropriate conditions, it is possible to accurately extract a unique cipher key for use with standard encryption algorithms. The experimental result is however different with real biometrics. Behavioural biometrics (such as the signature) showed very high variation in the measured features, and thus the corresponding FRR, are likely to be significant as demonstrated in their worst case scenario investigations. [4, 46] also identified one of the new challenges in using biometric to generate key in cryptography as the generation of unstable encryption key. [45] proposed a cryptographic key generation technique that made use of finger vein pattern. Finger vein is one of the newest biometric method which is more accurate than the traditional fingerprint. However, the authors did not carry out an assessment of the quality properties of the generated keys.

# 3. DEDUCTIONS AND POSSIBLE SOLUTIONS

Existing standard methods to ensuring privacy and security are characterised by utilization of cryptography which involves enciphering of patient data (Hewitt, 2013). In the existing models:

- Patients' awareness has been seen as an important factor to managing EHR. Autonomous Patient Controlled Encryption (PCE) and patients controlled privacy and security have been proposed [28][7]. There are important challenges in implementing personally controlled systems on a large scale because no matter how well these are integrated with institutional information systems, it is unlikely that patient controlled records could entirely replace provider or hospital based records.

- Private and public key cryptography methods have been widely proposed as the encryption mechanism of choice for EHR with difficulty in key management [7] [41] [27][16]. The Achilles' heel of cryptography approaches is the secrecy of encryption keys. Once any of key storage, generation or sharing schemes is breached, cryptography technique becomes void.

- There is the challenge of not providing patients' right and how the system can pinpoint the person who broadcast medical records for accountability responsibility [1]. In some cases where accountability have been provided, it only works during emergency access or by exposing physicians data [41] [29] [14]. Records have shown that EHRs have been breached outside emergency access [48].

- Lastly, no work is yet known of, at the moment, which uses bio cryptography technologies in the e-Health EHR domain and search is still in progress to for such. For EHR to be widely adopted, a strong and usable access control mechanism should be put in place to increase the patients' trustworthiness of their health data management system.

Considering these limitations of standard models of EHR systems from the physician's end, the patient and at EHR system's end, the following fundamental security issues are pertinent:

- How can sharable keys be generated?

- How can cryptographic keys be managed and secured?

- How can a data subject determine the amount of information available to requesters of health records? i.e. Issue of privacy

- Could access to EHR be audited accurately? Issue of accountability.

These challenges in existing cryptographic models alone in EHRs will contribute to the exploration of this study, For EHRs to be able to work without patients' fear of insecurity of health data stored by the system, it is pertinent that the three concepts of *privacy, security* and *accountability* be implemented individually and integrated into a single improved system. The future research should propose and implement a scheme that will nest each of the three as shown in Figure 1.

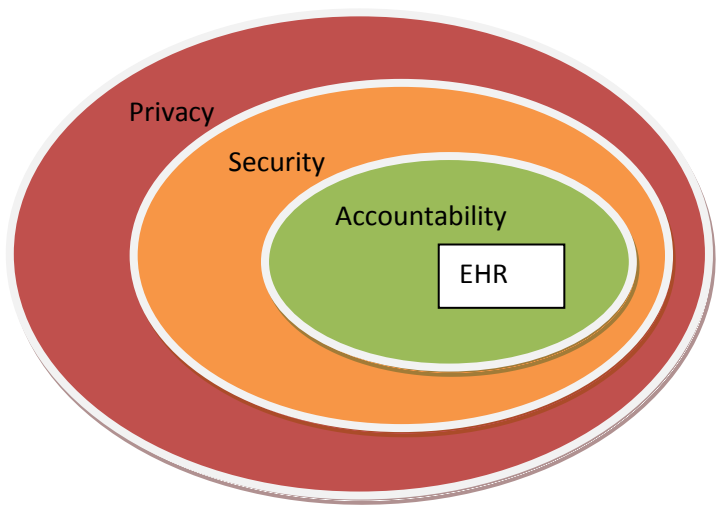

**Figure 1: proposed order of EHR implementation**

A pragmatic filter approach to control access rights should be implemented and should be jointly manageable by both the patients and the physicians. Privacy scheme devised this way will ensure that data fetched for every requester is based on sharing policy previously established.

Security via encryption would ensure that only the legitimate users can access records and as established from literature, this will further limit who can access available data in plain form. Furthermore with respect to cryptographic keys, bio cryptography approach could greatly solve the problem in key binding mode rather than key generation. To share a stable key with a physician without physical presence, a different method for generating stable and sharable key must be developed.

Last, upon being granted access, accountability should be implemented to ensure that parties accessing EHR of patients cannot repudiate operations performed.

## 4. CONCLUSION

As this is a part of an ongoing work, the completed research works is expected to propose and implement solutions to some of the identified challenges in implementing privacy, security, accountability and key management in electronic health record technology.